\documentclass[%
reprint,
showpacs,preprintnumbers,
 amsmath,amssymb,
 aps,
prl,
]{revtex4-1}

\usepackage{graphicx}% Include figure files
\usepackage{dcolumn}% Align table columns on decimal point
\usepackage{bm}% bold math
\begin{document}

%\preprint{APS}

\title{Stationary Large Amplitude Dynamics of the Finite Chain of Harmonically Coupled Pendulums 
}% Force line breaks with \\

\author{Valeri V. Smirnov}
\email{vvs@polymer.chph.ras.ru}
\author{Leonid I. Manevitch}%
 
\affiliation{%
 Institute of Chemical Physics, RAS, Moscow, Russia\\
119991, 4 Kosygin str., Moscow, Russia\\
}%
\date{\today}% It is always \today, today,

\begin{abstract}
We present an analytical description of the large-amplitude stationary oscillations of the finite discrete system of harmonically-coupled pendulums without any restrictions to their amplitudes (excluding   a  vicinity of $\pi$). 
Although this model has numerous applications in different fields of physics it was studied earlier in the infinite limit only. 
The developed approach allows to find the dispersion relations for arbitrary amplitudes of the nonlinear normal modes. 
We underline that the long-wavelength approximation, which is described by  well- known sine-Gordon equation leads to inadequate zone structure for the amplitude order of $\pi/2$ even if the chain is long enough. 
The extremely complex zone structure at the large amplitudes corresponds to  lot of resonances between nonlinear normal modes even with strongly different wave numbers.
Due to complexity of the dispersion relations the more short wavelength modes can possess the smaller frequencies. 
The numerical simulation of the dynamics of the finite-length chain is in a good agreement with obtained analytical predictions.
\end{abstract}

\maketitle

\section{Introduction}\label{s_I}

The wide class of physical models, the most known of which is the Frenkel-Kontorova (FK) one\citep{Frenkel&Kontorova1938, Braun&Kivshar2004} is based on the dynamics of the pendulum.
%The pendulum as a physical object has the long-time history of the study.
%The dynamics of the pendulum consists the base of the wide set of the physical models the most known of which is the Frenkel-Kontorova (FK) model \citep{Frenkel&Kontorova1938, Braun&Kivshar2004}.
The physical applications of the FK model comprise the theory of  dislocations \citep{Frenkel&Kontorova1938} and the problem of self-interstitial defects /crowdions/ in the crystal lattices \citep{Paneth1950}, in the problem of adsorption of a sub-monolayer films of atoms on a crystal surface  \citep{ Lyuksyutov1988},the theory of Josephson junctions \citep{ Hontsu&Ishii1988}, the theory of the proton conductivity of hydrogen-bonded chains \citep{Antonchenko1983}, in the description of the dynamics of planar defects such as twin boundaries \citep{Suezawa&Sumino1976} and domain walls in ferroelectrics \citep{Doring1948}, and ferro- or antiferromagnetics \citep{Enz1964},  certain biological processes like the DNA dynamics and denaturation \citep{Yomosa1983} (more references may be found in \citep{Braun&Kivshar2004}).
The common peculiarity of the mentioned models is the presence of the periodic on-site potential, while the inter-particle interaction can be described by potentials with a nonlinearity of different types.
The lucky star of  the Frenkel-Kontorova model is the existence of the integrable continuum limit of the respective equation of motion (sin-Gordon equation).
Due to full integrability of the latter its spectra of the nonlinear periodic  and localized excitations have been studied in detail \citep{Zakharov}.
%However, the continuum limit of the FK model with nonlinear interatomic potentials does not lead to integrable equations and the respective studies can be made by the approximate or numerical methods. 
The continuum limit of the discrete model with the nonlinear periodic inter-atomic interaction leads to the same sine-Gordon equation with understandable restrictions to the wavelengths (accounting the discreteness effects in the framework of approach introducing by Rosenau \citep{Rosenau1986} leads to  improving  the long-wavelength approximation only). 
Periodic interatomic potentials arise, in particular, while dealing with magnetic systems, unzipping the DNA molecule and oscillations of the flexi-chain polymers (see, e.g. \citep{Takeno1986, Takeno&Peyrard1996, Takeno&Peyrard1997}, where the existence of the highly localized soliton-like solution has been proved).
The main goal of this Letter is the study of nonlinear normal modes (NNMs) of harmonically coupled finite lattice in the wide range of the oscillation amplitudes and wavelengths. 
We propose a new semi-inverse asymptotic approach, which was successfully verified for two coupled pendulums in \citep{Manevitch2015}.
 This approach has allowed to reveal the extremely complex zone structure at the large amplitudes that leads to lot of resonances between NNMs with strongly different wave numbers.

\section{The model}\label{SLmodel}

We  consider a finite chain of the coupled particles with periodic on-site and inter-particle potentials; each of them is described by the harmonic function with, generally speaking, different periods.
This system will be referred as Sine-Lattice (SL) in contrast to the classic FK system.
Keeping in mind the coincidence of the mathematical descriptions we will discuss all results in the terms of coupled pendulums.
The periodic boundary conditions are the most appropriate  for the analysis of the chain dynamics.
The energy of such system may be written as follows:
\begin{equation}\label{eq:SLenergy}
\begin{split}
H=\sum_{j=1}^{N} \Bigl [\frac{1}{2} \left(\frac{d q_{j}}{d t}\right)^{2}+& \frac{\beta}{\alpha^{2}} \left(1-\cos {\left(\alpha (q_{j+1}-q_{j} ) \right)} \right)  \\
+&\left(1-\cos{q_{j}} \right) \Bigr]; \quad j=1,\dots ,N 
\end{split}
\end{equation}

where $q_{j}$ is the deviation of the $j-$th pendulum, while $\beta$ and $\alpha$ are the parameters, which specify the rigidity and the period of inter-pendulum coupling.
According to the periodic boundary conditions we assume that $q_{N+1}=q_{1}$ and $q_{0}=q_{N}$.

The respective equations of motion can be written as follows 
\begin{equation}\label{eq:SL0}
\begin{split}
\frac{d^{2}q_{j}}{d t^{2}} + & \frac{ \beta}{\alpha} \bigl ( \sin{(\alpha(q_{j+1}- q_{j}))}  \\
 - & \sin{(\alpha (q_{j}-q_{j-1}))} \bigr)+\sin{q_{j}}=0.
\end{split}
\end{equation}

%where $q_{j}$ is the angular coordinate of the j-th pendulum, $\tau_{0} =\omega_{0}t$, $\omega_{0}$ being its linear natural frequency, and $\varepsilon <<1$. As known, these equations describe also a basic approximation for description Josephson junction [3] as well as a particular case of the Frenkel-Kontorova model, having numerous applications in solid state physics and photonics [2, 6]. 

%\begin{multline}\label{eq:SG1}
%\frac{d^{2}q_{j}}{d \tau_{0}^{2}}+\omega^{2} q_{j}+\varepsilon \beta (q_{j+1}-2 q_{j}+q_{j-1})+  \\  \mu(\sin{q_{j}}-\omega^{2}q_{j})=0, \quad j=1,\ldots, N
%\end{multline}

%where $0 < \omega \leq  1$ (the lower limit will be later on speciﬁed) is the resonance oscillation frequency. Under internal 1:1 resonance conditions the combination of two terms in the second brackets has to be small (we suppose of order-$\varepsilon$), and $\mu = \varepsilon^{-1}$ is a book keeping parameter. Thus, we assume the closeness to resonance but we do not impose any restriction to the oscillations amplitude and the ensuing resonance frequency. 

Introducing the complex variables  given by 
\begin{equation}\label{eq:complex}
\begin{split}
\Psi_{j} = & \frac{1}{\sqrt{2}}(\frac{1}{\sqrt{\omega}} \frac{d q_{j}}{d t}+ i \sqrt{\omega} q_{j})  \\
q_{j} = & \frac{-i}{\sqrt{2 \omega}}(\Psi_{j}-\Psi_{j}^{*}),   \quad
\frac{d q_{j}}{d t} =  \sqrt{\frac{\omega}{2}}(\Psi_{j}+\Psi_{j}^{*})
\end{split}
\end{equation}

and substituting those in \eqref{eq:SL0}, one can rewrite equation \eqref{eq:SL0} as  
%\begin{widetext}
\begin{equation}\label{eq:SL1}
\begin{aligned}
 i \frac{d \Psi _{j}}{d t}+\frac{\omega}{2}   \left(\Psi _{j}+cc \right)   \\
 +  \frac{1}{\sqrt{2\omega}} \sum_{k=0}^{\infty}  \frac{1}{(2k+1)!} \left(\frac{1}{2 \omega} \right)^{k} \Bigl[ \left( \Psi_{j}-cc \right)^{2k+1}   \\
 -  \beta \, \alpha^{2k}  \bigl( ( \Psi_{j+1}-\Psi_{j}  - cc)^{2k+1}  -  \\   \left( \Psi_{j}-\Psi_{j-1}-cc \right)^{2k+1} \bigr) \Bigr]  =0,
\end{aligned}
\end{equation}
%\end{widetext}

%& i \frac{d \Psi _{j}}{d t}+\frac{\omega}{2}   \left(\Psi _{j}+\Psi _{j}^{*} \right)  \\ 
 %&+  \frac{1}{\sqrt{2\omega}} \sum_{k=0}^{\infty}{\frac{1}{(2k+1)!}\frac{1}{\sqrt{2 \omega}} \left( \Psi_{j}-\Psi_{j}^{*} \right)}   \\
%& -  \frac{\beta}{\alpha \sqrt{2\omega}} \sum_{k=0}^{\infty}\frac{1}{(2k+1)!}\left( \frac{\alpha}{\sqrt{2 \omega}} \right)^{2k+1} \Bigl[ ( \Psi_{j+1}-\Psi_{j}   \\
%& -  \Psi_{j+1}^{*}+\Psi_{j}^{*} )^{2k+1}  -    \left( \Psi_{j}-\Psi_{j-1}-\Psi_{j}^{*}+\Psi_{j-1}^{*} \right)^{2k+1} \Bigr]  =0

where the nonlinear terms in equation \eqref{eq:SL0} are represented as  the series of their arguments and abbreviation "cc" corresponds to the complex conjugated functions.

The efficient semi-inverse procedure for dynamic analysis without any restrictions to the oscillation amplitudes assumes that the considered system admits two time scales (fast and slow). 
Corresponding small parameter as well as the frequency  $\omega$ are not present in starting equations of motion \eqref{eq:SL0} and have to be determined later.
The solution of equation \eqref{eq:SL1} can be represented  as follows:
\begin{equation}\label{eq:SLsolution}
\Psi_{j}=\varphi_{j} e^{i \omega t}
\end{equation}

where $\varphi_{j}$ is a slow-changing function (the function of a "slow" time $\tau$).
%In such a case the extraction of the secular terms of equation \eqref{eq:SL1} leads to the equations for the functions $\varphi_{j}$:
After substituting \eqref{eq:SLsolution} into equations \eqref{eq:SL1}, integrating with respect to fast time and  providing absence of the secular terms, one obtains the equations for the functions $\varphi_{j}$:
%\begin{widetext}
\begin{equation}\label{eq:SLphi}
\begin{split}
i \frac{\partial \varphi_{j}}{\partial \tau}-\frac{\omega}{2} \varphi_{j}+  \frac{1}{\sqrt{2 \omega}}  J_{1} \left( \sqrt{\frac{2}{\omega }} \,  | \varphi_{j}| \right) \frac{\varphi_{j}}{| \varphi_{j} | } \\-\frac{\beta}{ \alpha\sqrt{2 \omega} }  \Bigl[ J_{1} \left( \alpha \sqrt{\frac{2}{\omega}} \, | \varphi _{j+1}- \varphi_{j} | \right) \frac{ \varphi _{j+1}- \varphi_{j} }{ | \varphi _{j+1}- \varphi_{j} | } \\
 - J_{1} \left( \alpha \sqrt{\frac{2}{\omega}} \, | \varphi _{j}- \varphi_{j-1}| \right) \frac{ \varphi_{j}- \varphi_{j-1}}{| \varphi_{j}- \varphi_{j-1} | }  \Bigr] =0
\end{split}
\end{equation}
%\end{widetext}

where $J_{1}$ is the  Bessel function of the first order.

In spite of complexity of equation \eqref{eq:SLphi}, one can directly check that the simple expression 
\begin{equation}\label{eq:SLsolution2}
\varphi_{j}= \sqrt{X} e^{- i \kappa j}
\end{equation}
with the wave number $\kappa=2 \pi /N$ satisfies it, if the frequency  $\omega$ is the solution of the equation
\begin{equation}\label{eq:SLX}
\begin{split}
 -\frac{\omega}{2} +  \frac{1}{\sqrt{2 \omega X}}  \Bigl[ 2\frac{\beta}{ \alpha } &   J_{1} \left( 2 \alpha \sqrt{\frac{2}{\omega}X} \sin{\frac{\kappa}{2}} \right) \sin{\frac{\kappa}{2}}  \\
 & + J_{1} \left( \sqrt{\frac{2}{\omega }X} \right)  \Bigr] =0
\end{split}
\end{equation}

Taking into account the relationship between the amplitude $X$ and  real amplitude of pendulum oscillations $Q$, which follows from definition \eqref{eq:complex}
\begin{equation*}
X=\frac{\omega}{2}Q^{2}
\end{equation*}

the transcendental equation \eqref{eq:SLX} is converted into extremely simple expression for  the NNMs frequencies
\begin{equation}\label{eq:eigenvalue}
\omega^{2}=\frac{2}{Q} \left(2 \, \frac{\beta}{\alpha} J_{1} \left(2 \, \alpha \, Q \sin{\frac{\kappa}{2}} \right) \sin{ \frac{\kappa}{2}}+J_{1} \left(Q \right) \right).
\end{equation}

Before study of  eigenfrequency \eqref{eq:eigenvalue} one should test the limit case, which corresponds to the oscillations of  a single pendulum.
Really, if the coupling parameter $\beta=0$,  hamiltonian  \eqref{eq:SLenergy} describes a set of independent pendulums, the oscillation frequency of which depends on the amplitude $Q$.
In such a case the frequency \eqref{eq:eigenvalue} has the form:
\begin{equation}\label{eq:single}
\omega=\sqrt{\frac{2}{Q} J_{1} (Q)}
\end{equation}

Equation \eqref{eq:single} can be compared with the exact oscillation frequency of pendulum:
\begin{equation}\label{eq:exact}
\omega_{e}=\frac{\pi \sin{\left( Q/2 \right) }}{2 F\left(Q/2,1/ \sin^2(Q/2) \right)}
\end{equation}

where $F$ is the elliptic integral of the first kind.

\begin{figure}
\includegraphics[width=50mm]{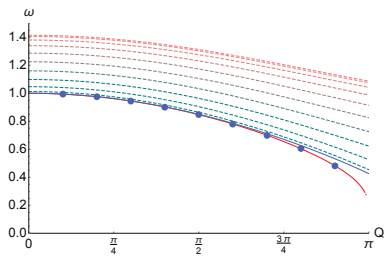}
\caption{(Color online) Comparison of pendulum's oscillation frequencies accordingly to eqs \eqref{eq:single} (black curve), \eqref{eq:exact} (red dot-dashed) and to numerical simulation (blue points).
The dashed curves show the zone structure for the FK chain with 20 pendulums. 
The coupling parameter $\beta=0.25$.}
\label{fig:singleP}
\end{figure}

%Figure \ref{fig:singleP} shows the comparison of oscillation frequencies described by equations \eqref{eq:single} and \eqref{eq:exact}. 
One can see from figure \ref{fig:singleP} that the agreement is excellent for all amplitudes  up to  $Q \simeq 3 \pi /4$.

%The reasons of the frequencies divergence is that when the oscillation amplitude $Q$  closes to maximum value $Q \simeq \pi$, the period of oscillation  enlarges and the procedure of the time separation (\ref{eq:SLsolution}, \ref{eq:SLphi}) turns out to be invalid.
The reason of the frequencies divergence is understandable: when the oscillation amplitude $Q$ approaches to maximum value $Q \sim  \pi$, the period of oscillation enlarges and the semi- inverse procedure (\ref{eq:SLsolution}, \ref{eq:SLphi}) is not self-consistent because of closing of two time scales.
 
Equation \eqref{eq:eigenvalue} describes the NNM "zone" structure, i.e. the dispersion ratio,  for the chain of harmonically coupled  pendulums at the arbitrary oscillation amplitude (excluding a vicinity of the "rotation limit" $Q=\pi$).

One should note that the "long wavelength" limit of equation \eqref{eq:eigenvalue} is not different from the classic FK model with a parabolic potential of inter-pendulum interactions $V \sim \beta (q_{j+1}-q_{j})^{2}$.
Really, considering the wave number $\kappa$ as a small value, one can expand the Bessel function into power series. 
The first term turns out to be $2 \alpha Q \sin{\kappa/2}$ and the respective eigenfrequency \eqref{eq:eigenvalue} is written as follows:
\begin{equation}\label{eq:FKeigenvalue}
\omega^{2}=2 \left(\frac{1}{Q} J_{1} \left(Q \right)  + 2 \, \beta \sin^{2}{\frac{\kappa}{2}}\right)
\end{equation}

Figure \ref{fig:singleP} shows the zone structure for the FK chain with 20 particles under periodic boundary conditions.

%\begin{figure}
%\includegraphics[width=50mm]{FKspectrum2}
%\caption{(Color online) The zone structure for the FK chain with 10 pendulums.  Solid curves correspond to exact (black) and present (blue) frequencies of uniform oscillations. The color dashed curves respect to the normal modes with various wave numbers. The coupling parameter $\beta=0.1$.}
%\label{fig:FKspectrum}
%\end{figure}

The low frequency zone bounding mode corresponds to the uniform oscillations of the chain or to the oscillations of the single pendulum \eqref{eq:single}, while the high-frequency bounding mode corresponds to the out-of-phase pendulum oscillations ("$\pi-$mode).
The zone width does not depend on the oscillation amplitude and the number of pendulums.
Only the gap value between different modes is the function of the length of the chain and the wave number.

\begin{figure}
\includegraphics[width=50mm]{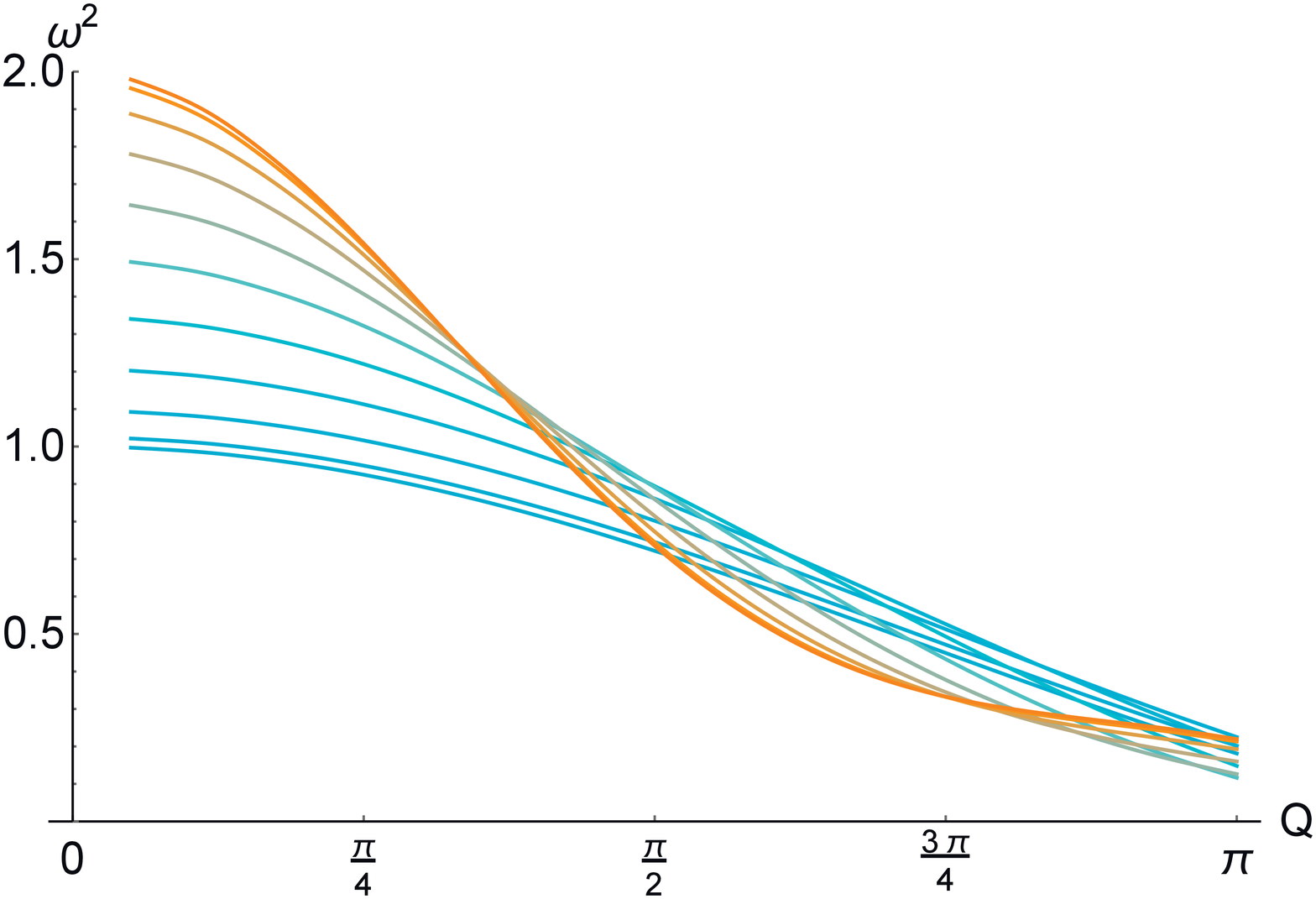}
\caption{(Color online) The zone structure for the SL chain with $100$ pendulums under periodic boundary conditions.  Light blue curve  and the light brown correspond to the zone bounding uniform and $\pi$-modes, respectively. Only each 5-th dispersion curve is shown in order to discern the separate curves. The coupling parameters $\beta=0.25$, $\alpha=1.2$.}
\label{fig:SLspectrum}
\end{figure}

Figure \ref{fig:SLspectrum} shows the zone structure for the harmonically coupled pendulums.

\begin{figure}
\includegraphics[width=50mm]{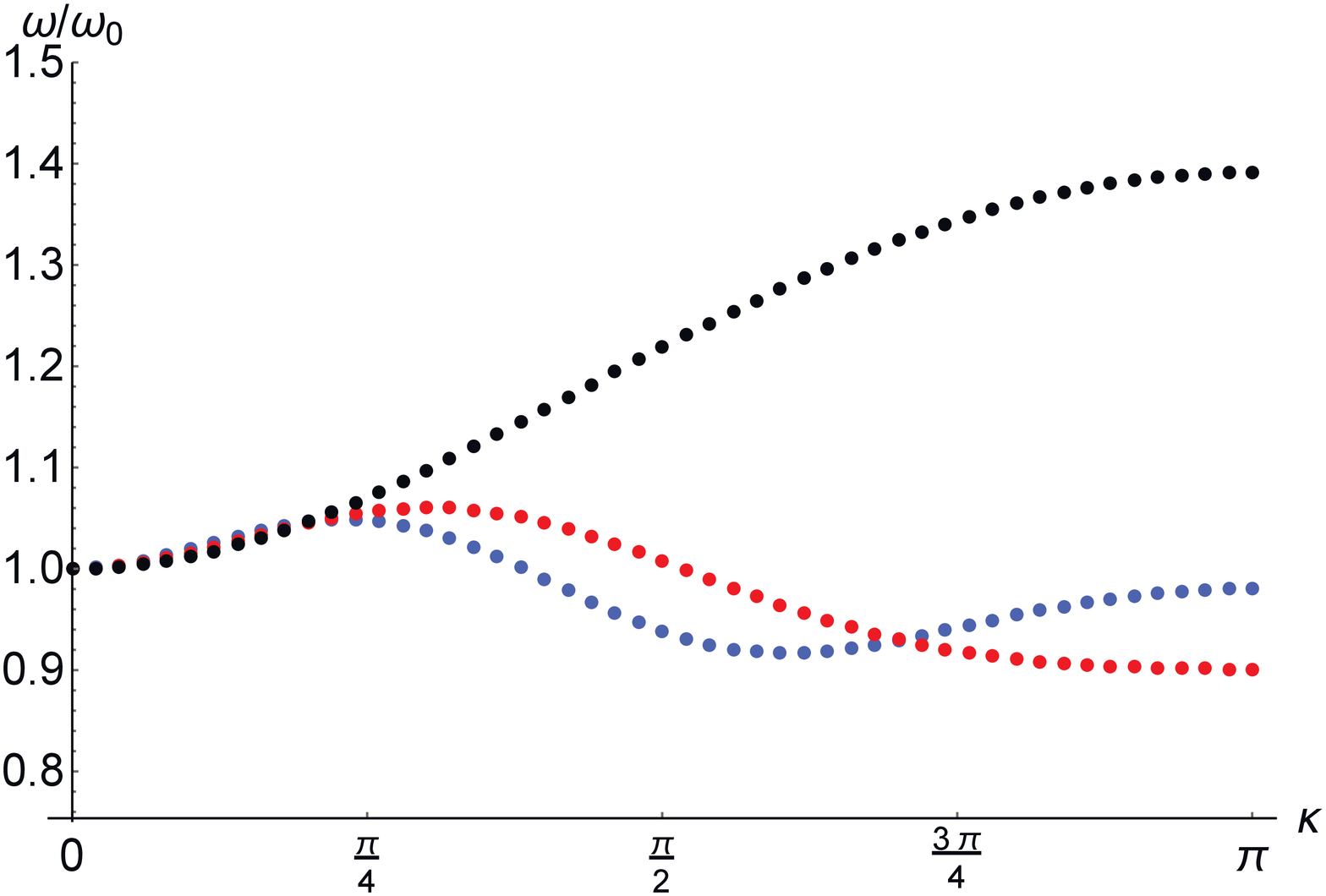}
\caption{(Color online) The comparison of the dispersion relations for the SL chain with 100 pendulums at the different oscillation amplitudes: black, red and blue points correspond to the amplitudes $Q=\pi/10$, $Q=7 \pi/10$ and $Q=9 \pi/10$, respectively. The relative frequencies $\omega / \omega_{0}$  ($ \omega_{0}$ is the frequency of the uniform modes) are shown. The potential parameters: $\beta=0.25$, $\alpha=1.2$.}
\label{fig:SLdispersion}
\end{figure}

The comparison of figures \ref{fig:singleP} and \ref{fig:SLspectrum} show a cardinal distinction between them.
Firstly, the width of the SL zone depends on the oscillation amplitude.
%What is more important: the NNMs with bigger wave numbers have the smaller frequencies when the amplitude value Q enlarges.
It is more important that the dispersion relation at a fixed amplitude $Q \geq \pi/2 $ (the threshold value depends on the parameters $\alpha$ and $\beta$) is a non-monotonic function of the wave number.
%But that is  more important some entanglement of the NNM frequencies, when the modes with a bigger wave number have the smaller frequencies than the modes with a smaller wave number, occurs when the amplitude value $Q$ enlarges.
As a result, the frequency of the zone bounding $\pi$-mode turns out to be smaller than the frequency of the uniform mode for large $Q$.
In such a case the multiple resonances occur in the vicinity of right edge of the spectrum, and the existence of them is defined by non-monotonic character of the dispersion relation rather than by the length of the chain.
Figure \ref{fig:SLdispersion} shows the dispersion relation for SL chain with 100 pendulums and the oscillation amplitude $Q=\pi/10$ in comparison with the same for $Q=9\pi/10$.

The structure of the NNMs zone for the SL chain has been checked by the direct numerical integration of equations \eqref{eq:SL0}.
\begin{figure}
\includegraphics[width=40mm]{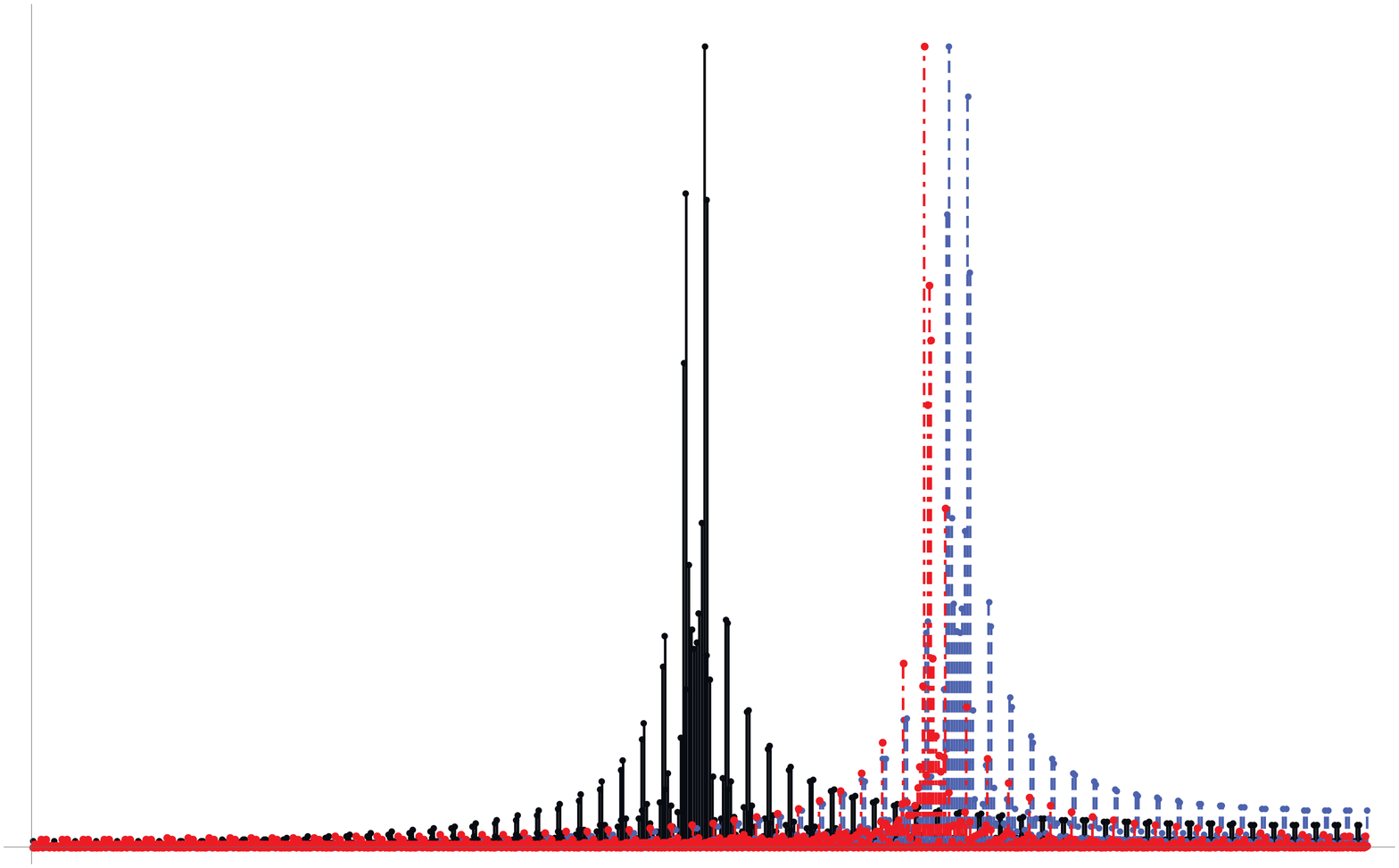}\quad \includegraphics[width=40mm]{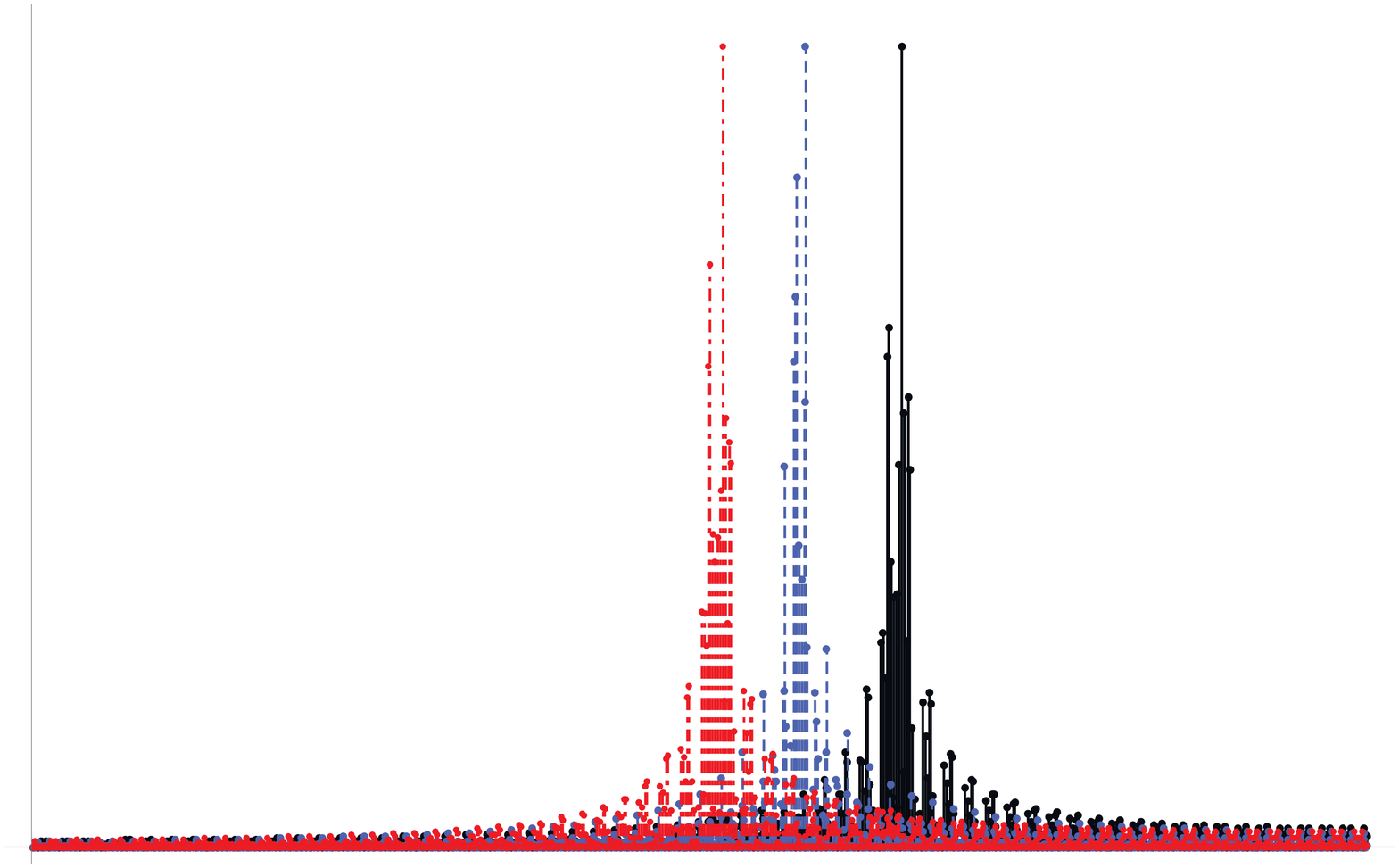}
\caption{(Color online) Fourier spectra of the oscillations of SL chain with 8 pendulums. Left panel shows the frequencies of the normal mode with $\kappa=0, \, 3\pi/4, \, \pi$ (solid black, dot-dashed red  and dashed blue lines, respectively) at the amplitude $Q=\pi/10$. Right panel - the same as left one at the amplitude $Q=9\pi/10$. The potential parameters: $\beta=0.25$, $\alpha=1.2$}
\label{fig:Fourier}
\end{figure}

Figure \ref{fig:Fourier} allows to compare the positions of the zone bounding modes and some intermediate one. 
One can see, that the mutual positions of the modes correspond to the dispersion relations that are shown in figure \ref{fig:SLdispersion}.

\section{Conclusion}\label{conclusion}

The study of normal modes of the SL chains revealed that  the high amplitude oscillations of harmonically coupled pendulums possess  non-trivial dispersion relation, when the normal modes with higher wave numbers correspond to smaller oscillation frequencies.
This peculiarity of the SL chain is not a consequence of the chain discreteness only, but is specified mainly by the type of the inter-pendulum potential. 
%This peculiarity of the SL chain is not an sequence of the chain discreteness, but is specified by the character of the inter-pendulum potential.
%It is important that this property is not reflected by the continuum limit of the dynamical equation, that is the well-known sin-Gordon one.
It is important that this property is not preserved in the continuum limit of the dynamical equation, which is the well-known sine-Gordon one.
The analytical semi-inverse approach with using the  multiple scale procedure  turns out to be very efficient for the investigation of  different types of nonlinear lattices. 
 In this connection we would like to refine the origin and the value of the small parameter, which allows to separate the time scales. 
 The natural small parameter in the problem considered above is the value of the gap between the frequencies of  neighbour NNMs. 
 As it follows from  equation \eqref{eq:eigenvalue} its value is provided by $\beta \sin^{2}{\kappa /2}$.
In spite of the fact that this parameter did not appear evidently, only its existence  allows us to develop equation \eqref{eq:SLphi}.
 In particular, the equations in the terms of the complex amplitudes, which are similar to equation \eqref{eq:SLphi}, allow to study the nonlinear normal mode interaction (see, e.g. \citep{Manevitch2010, Smirnov2010, Smirnov2011, Smirnov2014}), that opens a wide possibilities for the analysis of the nonlinear systems in the various fields of the physics.
 For example, the lattice with a wide class of nonlinear on-site as well as inter-site potentials may be easily analysed in the framework of this method.
 On the other side, the weakly coupled small systems may studied, if the coupling parameter of  is assumed to be small enough \citep{Manevitch2015}.

%-----------------------------------------------------------------

\begin{acknowledgments}
The work was supported by the Program of Department of Chemistry and Material Science (\#1), Russian Academy of Sciences, and the Russian Foundation for Basic Research (grant No 14-
01-00284a).
\end{acknowledgments}
  
\bibliography{SineLattice}% Produces the bibliography via BibTeX.

\begin{thebibliography}{20}%
\makeatletter
\providecommand \@ifxundefined [1]{%
 \@ifx{#1\undefined}
}%
\providecommand \@ifnum [1]{%
 \ifnum #1\expandafter \@firstoftwo
 \else \expandafter \@secondoftwo
 \fi
}%
\providecommand \@ifx [1]{%
 \ifx #1\expandafter \@firstoftwo
 \else \expandafter \@secondoftwo
 \fi
}%
\providecommand \natexlab [1]{#1}%
\providecommand \enquote  [1]{``#1''}%
\providecommand \bibnamefont  [1]{#1}%
\providecommand \bibfnamefont [1]{#1}%
\providecommand \citenamefont [1]{#1}%
\providecommand \href@noop [0]{\@secondoftwo}%
\providecommand \href [0]{\begingroup \@sanitize@url \@href}%
\providecommand \@href[1]{\@@startlink{#1}\@@href}%
\providecommand \@@href[1]{\endgroup#1\@@endlink}%
\providecommand \@sanitize@url [0]{\catcode `\\12\catcode `\$12\catcode
  `\&12\catcode `\#12\catcode `\^12\catcode `\_12\catcode `\%12\relax}%
\providecommand \@@startlink[1]{}%
\providecommand \@@endlink[0]{}%
\providecommand \url  [0]{\begingroup\@sanitize@url \@url }%
\providecommand \@url [1]{\endgroup\@href {#1}{\urlprefix }}%
\providecommand \urlprefix  [0]{URL }%
\providecommand \Eprint [0]{\href }%
\providecommand \doibase [0]{http://dx.doi.org/}%
\providecommand \selectlanguage [0]{\@gobble}%
\providecommand \bibinfo  [0]{\@secondoftwo}%
\providecommand \bibfield  [0]{\@secondoftwo}%
\providecommand \translation [1]{[#1]}%
\providecommand \BibitemOpen [0]{}%
\providecommand \bibitemStop [0]{}%
\providecommand \bibitemNoStop [0]{.\EOS\space}%
\providecommand \EOS [0]{\spacefactor3000\relax}%
\providecommand \BibitemShut  [1]{\csname bibitem#1\endcsname}%
\let\auto@bib@innerbib\@empty
%</preamble>
\bibitem [{\citenamefont {Frenkel}\ and\ \citenamefont
  {Kontorova}(1938)}]{Frenkel&Kontorova1938}%
  \BibitemOpen
  \bibfield  {author} {\bibinfo {author} {\bibfnamefont {Y.}~\bibnamefont
  {Frenkel}}\ and\ \bibinfo {author} {\bibfnamefont {T.}~\bibnamefont
  {Kontorova}},\ }\href@noop {} {\bibfield  {journal} {\bibinfo  {journal}
  {Phys. Z. Sowietunion}\ }\textbf {\bibinfo {volume} {13}},\ \bibinfo {pages}
  {1} (\bibinfo {year} {1938})}\BibitemShut {NoStop}%
\bibitem [{\citenamefont {Braun}\ and\ \citenamefont
  {Kivshar}(2004)}]{Braun&Kivshar2004}%
  \BibitemOpen
  \bibfield  {author} {\bibinfo {author} {\bibfnamefont {O.~M.}\ \bibnamefont
  {Braun}}\ and\ \bibinfo {author} {\bibfnamefont {Y.~S.}\ \bibnamefont
  {Kivshar}},\ }\href@noop {} {\emph {\bibinfo {title} {The Frenkel --
  Kontorova Model}}}\ (\bibinfo  {publisher} {Springer-Verlag},\ \bibinfo
  {address} {Berlin, Heidelberg},\ \bibinfo {year} {2004})\ p.\ \bibinfo
  {pages} {333}\BibitemShut {NoStop}%
\bibitem [{\citenamefont {Paneth}(1950)}]{Paneth1950}%
  \BibitemOpen
  \bibfield  {author} {\bibinfo {author} {\bibfnamefont {H.~R.}\ \bibnamefont
  {Paneth}},\ }\href@noop {} {\bibfield  {journal} {\bibinfo  {journal} {Phys.
  Rev.}\ }\textbf {\bibinfo {volume} {80}},\ \bibinfo {pages} {708} (\bibinfo
  {year} {1950})}\BibitemShut {NoStop}%
\bibitem [{\citenamefont {Lyuksyutov}\ \emph {et~al.}(1992)\citenamefont
  {Lyuksyutov}, \citenamefont {Naumovets},\ and\ \citenamefont
  {Pokrovsky}}]{Lyuksyutov1988}%
  \BibitemOpen
  \bibfield  {author} {\bibinfo {author} {\bibfnamefont {I.~F.}\ \bibnamefont
  {Lyuksyutov}}, \bibinfo {author} {\bibfnamefont {A.~G.}\ \bibnamefont
  {Naumovets}}, \ and\ \bibinfo {author} {\bibfnamefont {V.~L.}\ \bibnamefont
  {Pokrovsky}},\ }\href@noop {} {\emph {\bibinfo {title} {Two-Dimensional
  Crystals}}}\ (\bibinfo  {publisher} {Academic Press},\ \bibinfo {address}
  {Boston},\ \bibinfo {year} {1992})\BibitemShut {NoStop}%
\bibitem [{\citenamefont {Hontsu}\ and\ \citenamefont
  {Ishii}(1988)}]{Hontsu&Ishii1988}%
  \BibitemOpen
  \bibfield  {author} {\bibinfo {author} {\bibfnamefont {S.}~\bibnamefont
  {Hontsu}}\ and\ \bibinfo {author} {\bibfnamefont {J.}~\bibnamefont {Ishii}},\
  }\href@noop {} {\bibfield  {journal} {\bibinfo  {journal} {J. Appl. Phys.}\
  }\textbf {\bibinfo {volume} {63}},\ \bibinfo {pages} {2021} (\bibinfo {year}
  {1988})}\BibitemShut {NoStop}%
\bibitem [{\citenamefont {Antonchenko}\ \emph {et~al.}(1983)\citenamefont
  {Antonchenko}, \citenamefont {Davydov},\ and\ \citenamefont
  {Zolotaryuk}}]{Antonchenko1983}%
  \BibitemOpen
  \bibfield  {author} {\bibinfo {author} {\bibfnamefont {V.~Y.}\ \bibnamefont
  {Antonchenko}}, \bibinfo {author} {\bibfnamefont {A.~S.}\ \bibnamefont
  {Davydov}}, \ and\ \bibinfo {author} {\bibfnamefont {A.~V.}\ \bibnamefont
  {Zolotaryuk}},\ }\href@noop {} {\bibfield  {journal} {\bibinfo  {journal}
  {Phys. Stat. Sol. (b)}\ }\textbf {\bibinfo {volume} {115}},\ \bibinfo {pages}
  {631} (\bibinfo {year} {1983})}\BibitemShut {NoStop}%
\bibitem [{\citenamefont {Suezawa}\ and\ \citenamefont
  {Sumino}(1976)}]{Suezawa&Sumino1976}%
  \BibitemOpen
  \bibfield  {author} {\bibinfo {author} {\bibfnamefont {M.}~\bibnamefont
  {Suezawa}}\ and\ \bibinfo {author} {\bibfnamefont {K.}~\bibnamefont
  {Sumino}},\ }\href@noop {} {\bibfield  {journal} {\bibinfo  {journal} {Phys.
  Stat. Sol. (a)}\ }\textbf {\bibinfo {volume} {36}},\ \bibinfo {pages} {263}
  (\bibinfo {year} {1976})}\BibitemShut {NoStop}%
\bibitem [{\citenamefont {Doring}(1948)}]{Doring1948}%
  \BibitemOpen
  \bibfield  {author} {\bibinfo {author} {\bibfnamefont {W.}~\bibnamefont
  {Doring}},\ }\href@noop {} {\bibfield  {journal} {\bibinfo  {journal} {Z.
  Naturf.}\ }\textbf {\bibinfo {volume} {3a}},\ \bibinfo {pages} {373}
  (\bibinfo {year} {1948})}\BibitemShut {NoStop}%
\bibitem [{\citenamefont {Enz}(1964)}]{Enz1964}%
  \BibitemOpen
  \bibfield  {author} {\bibinfo {author} {\bibfnamefont {U.}~\bibnamefont
  {Enz}},\ }\href@noop {} {\bibfield  {journal} {\bibinfo  {journal} {Helv.
  Phys. Acta}\ }\textbf {\bibinfo {volume} {37}},\ \bibinfo {pages} {245}
  (\bibinfo {year} {1964})}\BibitemShut {NoStop}%
\bibitem [{\citenamefont {Yomosa}(1983)}]{Yomosa1983}%
  \BibitemOpen
  \bibfield  {author} {\bibinfo {author} {\bibfnamefont {S.}~\bibnamefont
  {Yomosa}},\ }\href@noop {} {\bibfield  {journal} {\bibinfo  {journal} {Phys.
  Rev. A}\ }\textbf {\bibinfo {volume} {27}},\ \bibinfo {pages} {2120}
  (\bibinfo {year} {1983})}\BibitemShut {NoStop}%
\bibitem [{\citenamefont {Novikov}\ \emph {et~al.}(1984)\citenamefont
  {Novikov}, \citenamefont {Manakov}, \citenamefont {Pitaevskii},\ and\
  \citenamefont {Zakharov}}]{Zakharov}%
  \BibitemOpen
  \bibfield  {author} {\bibinfo {author} {\bibfnamefont {S.}~\bibnamefont
  {Novikov}}, \bibinfo {author} {\bibfnamefont {S.}~\bibnamefont {Manakov}},
  \bibinfo {author} {\bibfnamefont {L.}~\bibnamefont {Pitaevskii}}, \ and\
  \bibinfo {author} {\bibfnamefont {V.}~\bibnamefont {Zakharov}},\ }\href@noop
  {} {\emph {\bibinfo {title} {Theory of Solitons: The Inverse Scattering
  Method}}},\ Monographs in Contemporary Mathematics\ (\bibinfo {year} {1984})\
  p.\ \bibinfo {pages} {276}\BibitemShut {NoStop}%
\bibitem [{\citenamefont {Rosenau}(1986)}]{Rosenau1986}%
  \BibitemOpen
  \bibfield  {author} {\bibinfo {author} {\bibfnamefont {P.}~\bibnamefont
  {Rosenau}},\ }\href@noop {} {\bibfield  {journal} {\bibinfo  {journal} {Phys.
  Lett. A}\ }\textbf {\bibinfo {volume} {118}} (\bibinfo {year}
  {1986})}\BibitemShut {NoStop}%
\bibitem [{\citenamefont {Takeno}\ and\ \citenamefont
  {Homma}(1986)}]{Takeno1986}%
  \BibitemOpen
  \bibfield  {author} {\bibinfo {author} {\bibfnamefont {S.}~\bibnamefont
  {Takeno}}\ and\ \bibinfo {author} {\bibfnamefont {S.}~\bibnamefont {Homma}},\
  }\href@noop {} {\bibfield  {journal} {\bibinfo  {journal} {J. Phys. Soc.
  Japan}\ }\textbf {\bibinfo {volume} {55}},\ \bibinfo {pages} {65} (\bibinfo
  {year} {1986})}\BibitemShut {NoStop}%
\bibitem [{\citenamefont {Takeno}\ and\ \citenamefont
  {Peyrard}(1996)}]{Takeno&Peyrard1996}%
  \BibitemOpen
  \bibfield  {author} {\bibinfo {author} {\bibfnamefont {S.}~\bibnamefont
  {Takeno}}\ and\ \bibinfo {author} {\bibfnamefont {M.}~\bibnamefont
  {Peyrard}},\ }\href@noop {} {\bibfield  {journal} {\bibinfo  {journal}
  {Physica D}\ }\textbf {\bibinfo {volume} {92}},\ \bibinfo {pages} {140}
  (\bibinfo {year} {1996})}\BibitemShut {NoStop}%
\bibitem [{\citenamefont {Takeno}\ and\ \citenamefont
  {Peyrard}(1997)}]{Takeno&Peyrard1997}%
  \BibitemOpen
  \bibfield  {author} {\bibinfo {author} {\bibfnamefont {S.}~\bibnamefont
  {Takeno}}\ and\ \bibinfo {author} {\bibfnamefont {M.}~\bibnamefont
  {Peyrard}},\ }\href@noop {} {\bibfield  {journal} {\bibinfo  {journal} {Phys.
  Rev. E}\ }\textbf {\bibinfo {volume} {55}},\ \bibinfo {pages} {1922}
  (\bibinfo {year} {1997})}\BibitemShut {NoStop}%
\bibitem [{\citenamefont {Manevitch}\ and\ \citenamefont
  {Romeo}(2015)}]{Manevitch2015}%
  \BibitemOpen
  \bibfield  {author} {\bibinfo {author} {\bibfnamefont {L.~I.}\ \bibnamefont
  {Manevitch}}\ and\ \bibinfo {author} {\bibfnamefont {F.}~\bibnamefont
  {Romeo}},\ }\href@noop {} {\bibfield  {journal} {\bibinfo  {journal}
  {Europhys. Lett.}\ } (\bibinfo {year} {2015})}\BibitemShut {NoStop}%
\bibitem [{\citenamefont {Manevitch}\ and\ \citenamefont
  {Smirnov}(2010{\natexlab{a}})}]{Manevitch2010}%
  \BibitemOpen
  \bibfield  {author} {\bibinfo {author} {\bibfnamefont {L.~I.}\ \bibnamefont
  {Manevitch}}\ and\ \bibinfo {author} {\bibfnamefont {V.~V.}\ \bibnamefont
  {Smirnov}},\ }\href@noop {} {\bibfield  {journal} {\bibinfo  {journal} {Phys.
  Rev. E}\ }\textbf {\bibinfo {volume} {82}},\ \bibinfo {pages} {036602}
  (\bibinfo {year} {2010}{\natexlab{a}})}\BibitemShut {NoStop}%
\bibitem [{\citenamefont {Manevitch}\ and\ \citenamefont
  {Smirnov}(2010{\natexlab{b}})}]{Smirnov2010}%
  \BibitemOpen
  \bibfield  {author} {\bibinfo {author} {\bibfnamefont {L.~I.}\ \bibnamefont
  {Manevitch}}\ and\ \bibinfo {author} {\bibfnamefont {V.~V.}\ \bibnamefont
  {Smirnov}},\ }\href@noop {} {\bibfield  {journal} {\bibinfo  {journal}
  {Doklady Physics}\ }\textbf {\bibinfo {volume} {55}},\ \bibinfo {pages} {324}
  (\bibinfo {year} {2010}{\natexlab{b}})}\BibitemShut {NoStop}%
\bibitem [{\citenamefont {Smirnov}\ and\ \citenamefont
  {Manevitch}(2011)}]{Smirnov2011}%
  \BibitemOpen
  \bibfield  {author} {\bibinfo {author} {\bibfnamefont {V.~V.}\ \bibnamefont
  {Smirnov}}\ and\ \bibinfo {author} {\bibfnamefont {L.~I.}\ \bibnamefont
  {Manevitch}},\ }\href@noop {} {\bibfield  {journal} {\bibinfo  {journal}
  {Acoustical Physics}\ }\textbf {\bibinfo {volume} {57}},\ \bibinfo {pages}
  {271 –} (\bibinfo {year} {2011})}\BibitemShut {NoStop}%
\bibitem [{\citenamefont {Smirnov}\ \emph {et~al.}(2014)\citenamefont
  {Smirnov}, \citenamefont {Shepelev},\ and\ \citenamefont
  {Manevitch}}]{Smirnov2014}%
  \BibitemOpen
  \bibfield  {author} {\bibinfo {author} {\bibfnamefont {V.~V.}\ \bibnamefont
  {Smirnov}}, \bibinfo {author} {\bibfnamefont {D.~S.}\ \bibnamefont
  {Shepelev}}, \ and\ \bibinfo {author} {\bibfnamefont {L.~I.}\ \bibnamefont
  {Manevitch}},\ }\href@noop {} {\bibfield  {journal} {\bibinfo  {journal}
  {Phys. Rev. Lett.}\ }\textbf {\bibinfo {volume} {113}},\ \bibinfo {pages}
  {135502} (\bibinfo {year} {2014})}\BibitemShut {NoStop}%
\end{thebibliography}%

\end{document}